\pdfoutput=1

\documentclass[11pt,twoside,a4paper,cmspaper,final,collab]{cms-tdr}
\def\svnVersion{337155}\def\cmsCernNoTag{CERN-PH-EP/2013-037}\def\cmsCernDate{\today}\def\cmsMessage{Published in Physics Letters B as \href{http://dx.doi.org/10.1016/j.physletb.2016.03.039}{\doi{10.1016/j.physletb.2016.03.039.}}}

\begin{document}\cmsNoteHeader{SUS-14-013}

\hyphenation{had-ron-i-za-tion}
\hyphenation{cal-or-i-me-ter}
\hyphenation{de-vices}
\RCS$Revision: 326070 $
\RCS$HeadURL: svn+ssh://svn.cern.ch/reps/tdr2/papers/SUS-14-013/trunk/SUS-14-013.tex $
\RCS$Id: SUS-14-013.tex 326070 2016-02-17 02:47:20Z paulini $
\newlength\cmsFigWidth
\ifthenelse{\boolean{cms@external}}{\setlength\cmsFigWidth{0.98\columnwidth}}{\setlength\cmsFigWidth{0.6\textwidth}}
\ifthenelse{\boolean{cms@external}}{\providecommand{\cmsLeft}{top}}{\providecommand{\cmsLeft}{left}}
\ifthenelse{\boolean{cms@external}}{\providecommand{\cmsRight}{bottom}}{\providecommand{\cmsRight}{right}}

\newcommand{\ttgamma}{{\ttbar\hspace{-2pt}\Pgg}}

\cmsNoteHeader{SUS-14-013}
\title{Search for supersymmetry in events with a photon, a lepton,
       and missing transverse momentum
       in $\Pp\Pp$ collisions at $\sqrt{s}$ = 8\TeV }

\date{\today}

\abstract{
  A search for supersymmetry involving events with at least one
  photon, one electron or muon, and large missing transverse momentum has
  been performed by the CMS experiment. The data sample corresponds to
  an integrated luminosity of 19.7\fbinv of $\Pp\Pp$ collisions at
  $\sqrt{s}=8\TeV$, produced at the CERN LHC. No excess of events is
  observed beyond expectations from standard model processes.  The
  result of the search is interpreted in the context of a general model
  of gauge-mediated supersymmetry breaking, where the charged and
  neutral winos are the next-to-lightest supersymmetric
  particles. Within this model, winos with a mass up to 360\GeV are
  excluded at the 95\% confidence
  level. Two simplified models inspired by gauge-mediated
  supersymmetry breaking are also examined, and used
  to derive upper limits on the production cross sections
  of specific supersymmetric processes.
}

\hypersetup{%
pdfauthor={CMS Collaboration},%
pdftitle={Search for supersymmetry in events with a photon, a lepton, and missing transverse momentum in pp collisions at sqrt(s) = 8 TeV},%
pdfsubject={CMS},%
pdfkeywords={CMS, physics, supersymmetry}}

\maketitle
\newcommand{\MT}{\ensuremath{M_{\mathrm{T}}}\xspace}
\providecommand{\PSW}{\ensuremath{\mathrm{\widetilde{W}}}\xspace}

\section{Introduction}
\label{sec:introduction}

The extension of the standard model (SM) of particle physics through the
concept of
supersymmetry (SUSY)~\cite{Martin:1997ns}, which imposes a symmetry between fermions and
bosons, can offer a
solution to some of the issues not accommodated in the SM, such as the
existence of dark matter in the universe or the extreme fine tuning
required to control radiative corrections to the Higgs boson mass (hierarchy
problem)~\cite{Wess:1974tw,Cheng:2003ju,Appelquist:2000nn}. The
minimal supersymmetric standard model
(MSSM)~\cite{Fayet:1976et, Farrar:1978xj, Dimopoulos:1981zb} provides a
calculable framework with a fully known particle content, introducing a
superpartner for each SM particle.
For example, squarks, gluinos, and gravitinos are the SUSY partners of
quarks, gluons, and gravitons, respectively.
The MSSM has guided the search program for
physics beyond the SM at facilities such as the Fermilab Tevatron and CERN
LHC. Existing searches have not yet found
evidence for SUSY, but a large parameter space of the MSSM
remains to be explored.

Within the MSSM, scenarios based on gauge-mediated SUSY breaking
(GMSB)~\cite{Dine:1981za,Dimopoulos:1981au,Dine:1981gu,Dine:1982zb,
  Nappi:1982hm,AlvarezGaume:1981wy,Dine:1993yw,Dine:1994vc,Dine:1995ag,
  Baer:1996hx,Dimopoulos:1996yq} are of particular interest because of
their ability to naturally circumvent the so-called SUSY flavour
problem~\cite{Dimopoulos:1995ju}.  The framework of general gauge
mediation (GGM)~\cite{Meade:2008wd} offers a clear definition of GMSB
and establishes its key aspects.  For example, GMSB predicts the
gravitino (\PXXSG) to be the lightest supersymmetric particle (LSP). The
combination of this feature and the weakness of the coupling of \PXXSG
to other MSSM particles has specific consequences in collider
phenomenology.  Under the assumption that R-parity~\cite{Farrar:1978xj}
is conserved, SUSY particles are pair-produced at the LHC. Except for
direct LSP pair production, each SUSY particle initiates a decay chain
that yields the next-to-lightest supersymmetric particle
(NLSP). Branching fraction for the SUSY particle decay involving \PXXSG
is negligible except for the NLSP, leaving the decay of the NLSP to its
SM partner and the \PXXSG as effectively the only gravitino production
mechanism.  The gravitino escapes detection, leading to missing momentum
in the event. The signature of a GMSB signal is thus strongly dependent
on the identity of the NLSP. In most GMSB models, the NLSP is taken to
be a bino- or wino-like lightest neutralino, where a bino and wino are the
superpartners of the SM U(1) and SU(2) gauge fields, respectively.
Previous searches for a GMSB signal typically exploited the diphoton
signature~\cite{Chatrchyan:2012bba, Aad:2012zza, Aaltonen:2009tp,
  Abazov:2010us, Heister:2002vh, Abdallah:2003np, Achard:2003tx,
  Abbiendi:2005gc, Aktas:2004cc}, in which each of the two bino-like
neutralinos decays promptly into a photon and a gravitino. Similar
scenarios with nonprompt NLSP decays have also been
considered~\cite{Chatrchyan:2012jwg,Aad:2013oua}. Thus far, no evidence
for GMSB SUSY has been observed, resulting in upper limits on the
production cross sections given as a function of the SUSY particle
masses, the NLSP lifetime, or other model parameters.

\begin{figure}[bth]
  \centering
    \includegraphics[width=0.48\textwidth]{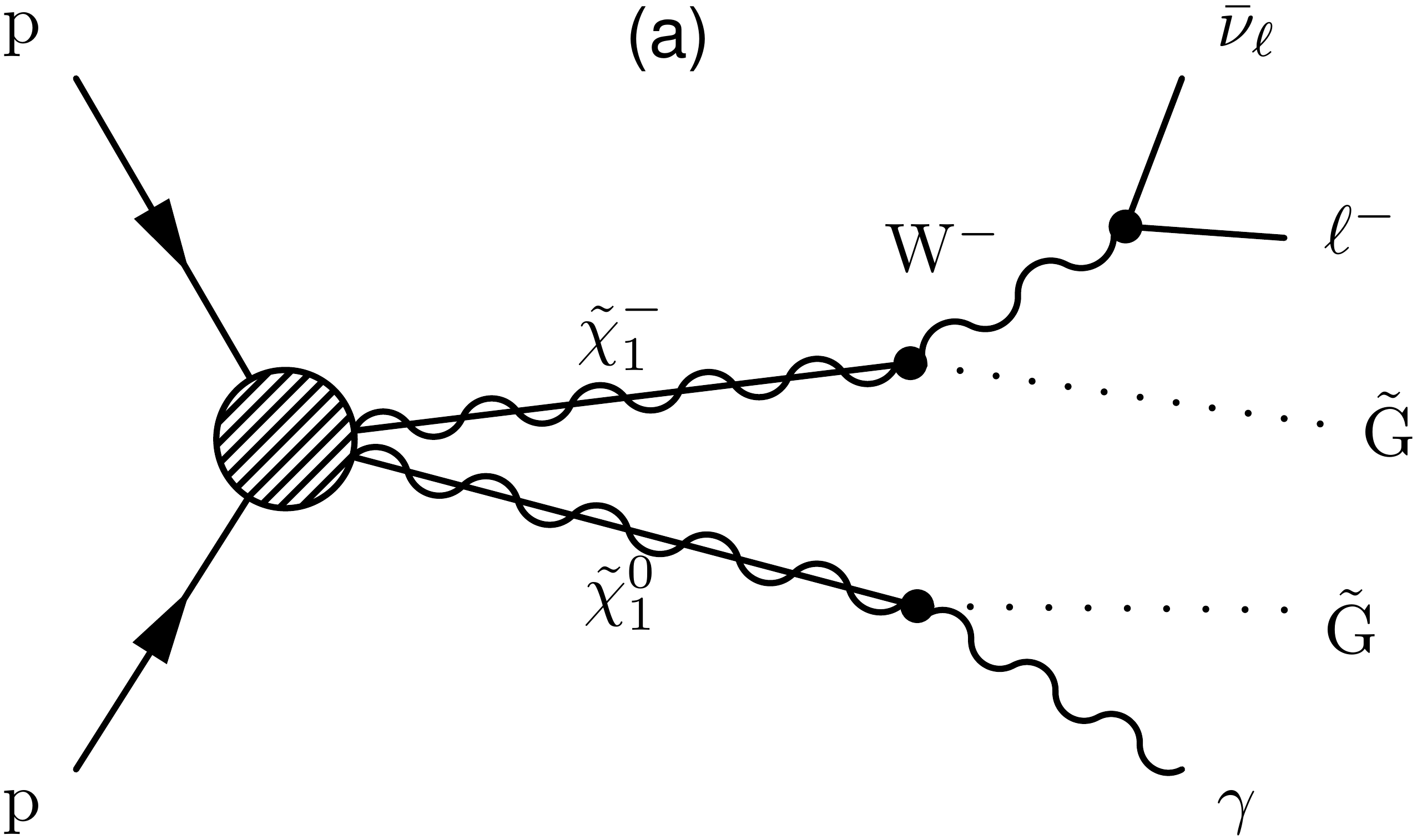}
    \includegraphics[width=0.48\textwidth]{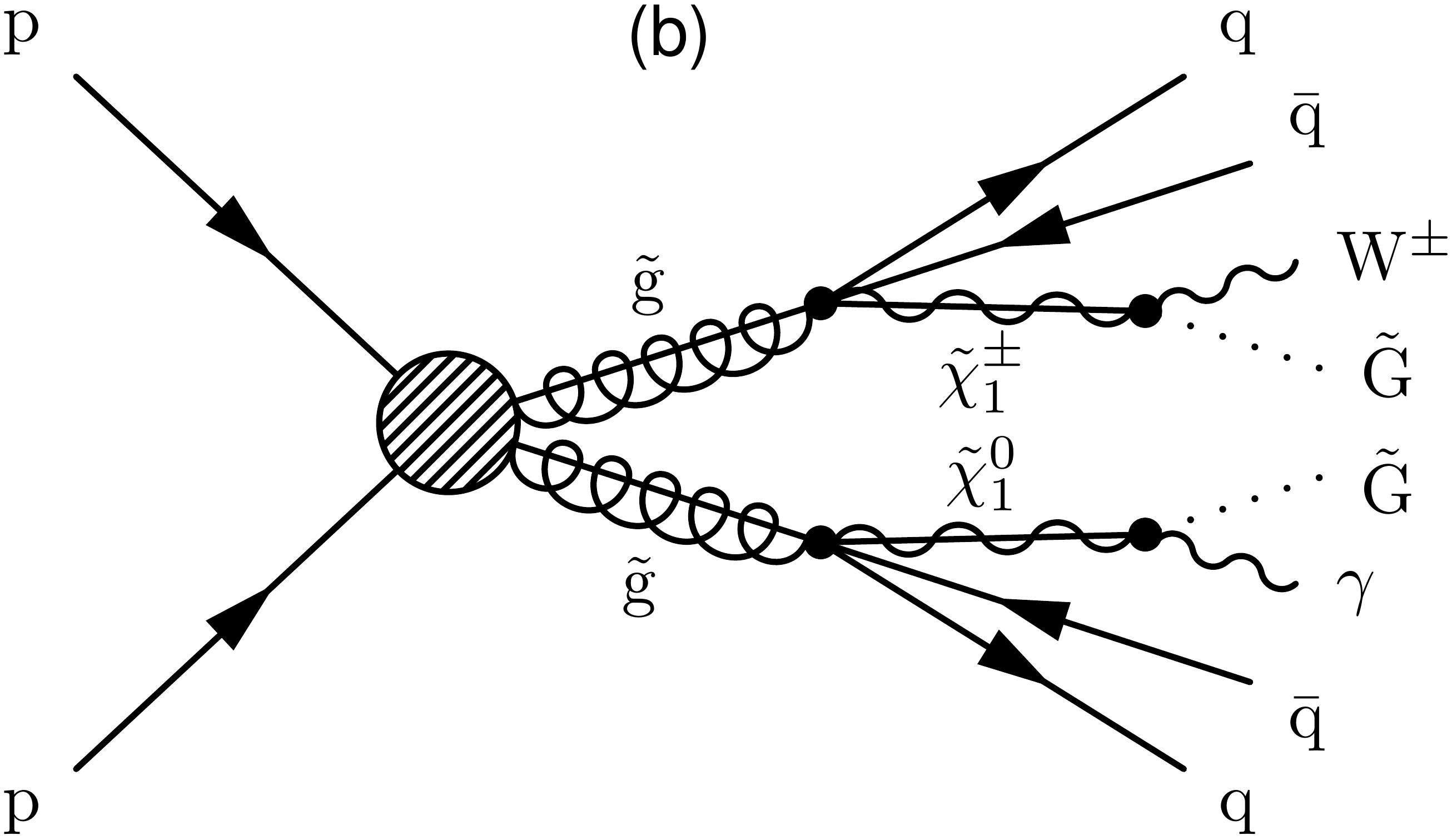}
    \caption{Diagrams showing the production and decays of wino-like
    co-NLSPs (\PSGcpmDo and \PSGczDo) leading to final states with a
    photon, an electron or muon, and missing momentum from undetected
    gravitinos \PXXSG (a) without and (b) with
    involvement of coloured SUSY particles.}
    \label{fig:SMS_cartoon}
\end{figure}

This paper presents a search for SUSY with the CMS experiment at the
LHC, and targets GGM models with wino-like NLSPs. The data sample
corresponds to an integrated luminosity of 19.7\fbinv of \Pp\Pp\
collision data collected in 2012 at $\sqrt{s}=8\TeV$. In particular, we
study the wino co-NLSP model~\cite{Ruderman:2011vv}, in which nearly
mass-degenerate charged and neutral winos are significantly lighter than
the other electroweakinos and constitute the lightest SUSY particles
aside from the gravitino.  Although the lifetime of the NLSP is
effectively a free parameter in GGM phenomenology, a prompt decay of
winos is assumed in this analysis.  A signature of at least one
photon~(\Pgg), one electron or muon~($\ell$), and large missing
transverse momentum~(\ptvecmiss) is used in this search. The photon is
assumed to be emitted by a neutralino NLSP, and the leptons by either a
charged or neutral NLSP decaying to a \PW\ or \cPZ\ boson,
respectively. This signature suppresses many SM backgrounds, obviating
the need for additional requirements such as associated jet
activity. The diagrams in Fig.~\ref{fig:SMS_cartoon} provide examples of
the decay chains studied in this analysis. The present search is
sensitive to the direct electroweakino production mode of
Fig.~\ref{fig:SMS_cartoon}\,(a), where the winos are produced without
involving coloured SUSY particles, but also to strong production modes
such as the gluino (\PSg) pair-production process shown in
Fig.~\ref{fig:SMS_cartoon}\,(b). Similar searches were conducted by the
ATLAS~\cite{Aad:2015hea} and
CMS~\cite{Chatrchyan:2011ah,Khachatryan:2015exa} experiments using LHC
pp collision data at $\sqrt{s}=7$ or 8\TeV, as well as the CDF
experiment~\cite{Acosta:2001ib} at the Tevatron using \Pp\Pap\ collision
data at $\sqrt{s}=1.8\TeV$. None of these analyses sees an excess of
events over the respective SM predictions.  The wino co-NLSP model has
also been probed through the signatures of three leptons or two leptons
and two jets~\cite{Aad:2014vma,Khachatryan:2014qwa}, which target the
decay of the neutralino NLSP to a gravitino and a \cPZ\ boson rather
than to a gravitino and a photon. None of these analyses observed a
significant excess of events over their respective SM predictions.

\section{CMS detector}
\label{sec:detector}

The central feature of the CMS apparatus is a superconducting solenoid
of 6\unit{m} internal diameter, providing a magnetic field of
3.8\unit{T}.  Within the solenoid volume are a silicon
pixel and strip tracker, a lead tungstate crystal electromagnetic
calorimeter (ECAL), and a brass and scintillator hadron calorimeter
(HCAL), each consisting of a barrel and two endcap sections. Muons are
measured in gas-ionization detectors embedded in the steel flux-return
yoke outside the solenoid. Extensive forward calorimetry complements the
coverage provided by the barrel and endcap detectors.  A detailed
description of the CMS detector, together with a definition of the
coordinate system and the relevant kinematic variables, can be found in
Ref.~\cite{Chatrchyan:2008aa}.

In the barrel section of the ECAL, an energy resolution of about 1\% is
achieved for unconverted and late-converting photons with
transverse energy $\ET\approx10\GeV$.
The remaining barrel photons have a resolution
of about 1.3\% up to a pseudorapidity $\abs{\eta} < 1.0$, rising to about
2.5\% for $\abs{\eta} = 1.4$~\cite{Khachatryan:2015iwa}.

The electron momentum is determined by combining the energy measurement
in the ECAL with the momentum measurement in the tracker. The momentum
resolution for electrons with transverse momentum $\pt\approx45\GeV$
from $\cPZ\to\Pep\Pem$ decays ranges from 1.7\% for non-showering
electrons in the barrel region to 4.5\% for showering electrons in the
endcaps~\cite{Khachatryan:2015hwa}.

Muons are measured in the range $\abs{\eta}< 2.4$, with
detector elements based on three technologies: drift tubes, cathode
strip chambers, and resistive plate chambers.
Through the matching of track segments measured in the muon detectors with
tracks measured in the tracker, a transverse momentum resolution
of 1.3--2.0\% is achieved for barrel muons with $20<\pt<100\GeV$.
In the endcaps, the resolution increases up to around 6\%.
The \pt resolution in
the barrel is better than 10\% for muons with transverse momentum up to
1\TeV~\cite{Chatrchyan:2012xi}.

Physics objects are defined using the particle-flow (PF) algorithm~\cite{CMS-PAS-PFT-09-001,CMS-PAS-PFT-10-001}, which
reconstructs and identifies individual particles through an
optimized combination of information from different elements of the CMS
detector. The PF candidates are classified as photons, charged hadrons,
neutral hadrons, electrons, or muons. Finally, the CMS detector is
nearly hermetic, permitting accurate measurements of \ptvecmiss.

\section{Data collection and event selection}
\label{sec:data_collection}

The search is
conducted in the electron-photon (\Pe\Pgg) and muon-photon (\Pgm\Pgg)
channels. The data samples are collected using a dedicated trigger for
each channel, as described below.
An event is considered to be in the
\Pe\Pgg\ (\Pgm\Pgg) channel if it contains at least one high-energy
photon and an electron (muon). Events that simultaneously satisfy the
criteria for the two search channels, representing about 0.1\% of the
selected events, are classified as \Pgm\Pgg\ candidates because muon
objects are less often the result of hadron misidentification than are
electron objects.

The trigger for the \Pe\Pgg\ channel requires at least two isolated
photon-like objects, with \ET\ thresholds of 36 and 22\GeV for
the highest and second-highest \ET\ photon, respectively. The trigger does not
veto photon objects that can be matched to a track, allowing events with a photon and an electron to
also satisfy the trigger. The \Pgm\Pgg\ channel uses a muon-photon
trigger with a \pt\ threshold of 22\GeV for both the photon and muon
objects. To ensure a fully efficient trigger and a similar
selection efficiency for the two channels, the subsequent analysis
requires $\ET > 40\GeV$ for the photon and $\pt > 25\GeV$ for the electron
or muon. With these requirements, the trigger efficiency for the
signal models described in Section~\ref{sec:interpretation} is found to be 93--98\% for
both channels, depending on the model and SUSY mass values.

Photon candidates are reconstructed from clusters of energy in the
ECAL~\cite{Khachatryan:2015iwa}. The momentum vector of the photon
points from the primary \Pp\Pp\ interaction vertex to the center of the
ECAL energy cluster, under the assumption that the photon originates from the
primary vertex, which is defined as the vertex with the highest $\sum
\PT^2$ of associated tracks.  Only photons from clusters in the
pseudorapidity range $\abs{\eta} < 1.44$ are included in this analysis.
These clusters were selected as photon candidates by a set of criteria that
are designed to achieve a 90\% identification efficiency for true photons. For
a cluster to be identified as a photon, its shape must be consistent
with that expected from a photon, and the energy detected in the HCAL
behind the cluster cannot exceed 5\% of the ECAL energy. To further
suppress the misidentification of hadrons as photons, a PF-based
isolation requirement is imposed. The transverse component of the
momentum sum of each of the PF
photons, charged hadrons, and neutral hadrons within a cone of $\Delta R
\equiv \sqrt{ \smash[b]{(\Delta\eta)^2+(\Delta\phi)^2}} = 0.3$ around the direction
of the photon candidate (where $\phi$ is the azimuth measured in
radians) is required not to exceed fixed values defined to achieve a
desirable balance between the identification efficiency and misidentification
rate. The photon object that is being identified is not included in the
isolation sums, and charged hadrons
are included only if they are associated with the primary
vertex. The \PT\ sums are corrected for contributions from
additional \Pp\Pp\ interactions (pileup). To distinguish photon
candidates from isolated electrons, photon objects are vetoed if a
matching track segment from the inner tracker is identified.

Electron (muon) candidates must lie in the pseudorapidity range
$\abs{\eta}<2.5\,(2.4)$. For electrons, the transition region
$1.44<\abs{\eta}<1.56$ between the barrel and the endcap detectors is
vetoed because the reconstruction efficiency in this region is difficult
to model. Electron objects are reconstructed by associating a cluster of
energy deposited in the ECAL with a reconstructed track. The electron
selection~\cite{Chatrchyan:2013dga} is based on the shower shape, the
matching of a track to the cluster, and isolation, where the isolation
variable is calculated from the momenta of PF photons, charged hadrons,
and neutral hadrons within a cone of $\Delta R = 0.3$ around the
electron direction, corrected for the effects of pileup. The isolation
sum is required not to exceed a fixed fraction of the electron \PT, where
the selection criteria are defined to obtain an 80\% electron
identification efficiency. The muon selection~\cite{Chatrchyan:2012xi},
targeting a 90\% efficiency for true muons, utilizes the quality of the
track fit, the number of detector hits used in the tracking, and the
isolation.  The isolation requirement for muons is similar to that for
electrons, but uses a larger cone size $\Delta R = 0.4$.
Electrons and muons must originate from a primary vertex, with
respective distances of closest approach for electrons (muons) of less
than 0.2\mm (2\mm) in the transverse plane and $<$1\mm ($<$5\mm) along the
beam direction.

The reconstruction of jets and \ptvecmiss is also based on the PF objects. All
reconstructed PF candidates are clustered into jets using the
anti-\kt clustering
algorithm~\cite{Cacciari:2008gp,Cacciari:2011ma}, with a distance
parameter of 0.5. Jet objects are used to calculate the
\HT variable, defined as the scalar \pt sum of jets. To be considered in the \HT sum, a jet must have a
calibrated and pileup-corrected~\cite{Chatrchyan:2011ds} \pt value
greater than 30\GeV, $\abs{\eta}<2.5$, and be consistent
with an origin at the primary
vertex~\cite{CMS-PAS-JME-13-005}. In addition, it must be no closer than
$\Delta R=0.5$ to the photon or lepton candidates. The
missing momentum \ptvecmiss is given by the negative of the vector \PT
sum of all PF objects, with jet-energy corrections applied. The
magnitude of \ptvecmiss is referred to as \MET.

To suppress the background from final-state radiation events with an on-shell
\PW\ (\cPZ) boson that decays to $\ell\Pgn\Pgg$
($\ell\ell\Pgg$), the highest-\ET photon in an event must be separated
by $\Delta R > 0.8$ from the highest-\pt electron or
muon. Additionally, for the \Pe\Pgg\ channel, the invariant mass
of the electron-photon system is required to differ by more than
10\GeV from the
nominal \cPZ\ boson mass~\cite{Agashe:2014kda}, to
reduce background from electrons misidentified as photons.

After applying the selection requirements described above, the obtained
event yields are compared to expectations from SM background processes.
The signal region of interest is defined by $\MET>120\GeV$ and
$\MT>100\GeV$, where transverse mass \MT is defined by
$\MT=\sqrt{\smash[b]{2\MET\PT^{\ell}[1-\cos\Delta\phi(\ell,\ptvecmiss)]}}$,
with $\PT^{\ell}$ the transverse momentum of the highest-\pt lepton and
$\Delta\phi(\ell, \ptvecmiss)$ the azimuthal angle between
the lepton and \ptvecmiss.
The \MT requirement reduces backgrounds from processes that
produce \PW\ bosons.
Table~\ref{tab:cutflow} shows the observed number of events at different
stages of the selection process.
Because of a higher selection efficiency for muons, after implementing
the selection requirements, the number of observed events in the
\Pgm\Pgg\ channel is larger than in the \Pe\Pgg\ channel.

\begin{table*}[tbp]
  \centering
    \topcaption{
      Summary of event selection requirements and observed number of events
      after applying the listed selection requirements in
      successive order. The symbol $m_{\Pe\Pgg}$ and $m_{\cPZ}$ denote
      the invariant mass of the electron-photon system and the nominal
      \cPZ\ boson mass, respectively.
    }
    \label{tab:cutflow}
    \begin{tabular}{lrr}
      \hline
      Selection requirement & \Pe\Pgg\ channel & \Pgm\Pgg\ channel \\
      \hline
      Trigger & $26\,733\,051$ & $19\,456\,571$ \\
      $\geq$1 accepted $\Pgg$ & $2\,718\,364$ & $243\,664$ \\
      $\geq$1 accepted $\ell$ ($\ell = \Pe, \Pgm$)& $70\,736$ & $32\,173$ \\
      $\Delta R(\Pgg, \ell) > 0.8$ & $68\,168$ & $30\,232$ \\
      $\abs{m_{\Pe\Pgg} - m_{\cPZ}} > 10\GeV$ & $29\,169$ & --- \\
      $\MET > 120\GeV, \MT > 100\GeV$ & $110$ & $152$ \\
      \hline
    \end{tabular}

\end{table*}

\section{Background estimation}
\label{sec:backgrounds}

Three sources of SM background are considered: misidentified photons,
misidentified leptons, and electroweak backgrounds.

\subsection{Misidentified-photon background}
\label{subsec:fake_photons}

The background from misidentified photons arises from
events in which a photon object does not correspond
to a genuine prompt photon.  The dominant background processes
in this category are Drell--Yan dielectron ($\Pq\Paq\to\Pgg^{*}\to\Pep\Pem$)
and W ($\to\ell\nu$)+jets production, in which an electron or jet, respectively,
is misidentified as a photon.  Minor contributions
arise from \ttbar events with leptonic top quark decays, for
both the \Pe\Pgg\ and \Pgm\Pgg\ channels.  Events with \ttbar
production also contribute to the background
if a jet is misidentified as a photon.
An electron can be misidentified as a photon if it fails to register
track seeds due to detector inefficiencies such as non-operational
sensors in the tracker. A jet can be misidentified as a photon
if a large fraction of its energy is carried by
mesons decaying to photons, such as $\Pgpz\to\Pgg\Pgg$.
These two types of background are estimated from
data using weighted control samples. The method proceeds in two
steps. First, a control sample enriched in particles that are prone
to be misidentified as photon candidates, \ie, electrons or neutral
hadrons, is selected by inverting certain criteria in the photon
identification, while keeping the other selection requirements identical
to those for signal
candidates. This control sample is called the proxy sample. The
second step is to determine the transfer factor $N^{\text{misid}} /
N^{\text{proxy}}$, where $N^{\text{misid}}$ is the estimate of the
number of misidentified events in the signal candidate sample
and $N^{\text{proxy}}$ is the number of events in the proxy sample. The
proxy sample is then scaled by the transfer factor. The definition of
the proxy sample is tuned to make its kinematic properties similar to
those of events with misidentified photons.
Thus, this two-step procedure takes the
set of misidentified events in a control region where the SUSY signal
contribution is expected to be negligible, \eg in events with small
\MET, and utilizes it to model the distribution of the misidentified
background for a given kinematic variable. In particular,
from the extrapolation of the observed events in the
control region, the method predicts an expectation for the number of
events and corresponding kinematic distribution in the signal region.  
A detailed validation
of this background estimation is performed by
applying the method to Monte Carlo (MC)
simulation samples, and comparing the outcome of this procedure to the
known generated MC content. Good agreement is found in all such tests.

The proxy sample for events with an electron-to-photon misidentification
is constructed by inverting the electron-seed veto in the photon
identification, which turns the photon candidate into an electron
proxy. The transfer factor for this proxy sample is determined
by counting $\cPZ\to\Pep\Pem$ decays in a separate control sample,
defined by $\MET < 70\GeV$. The ratio
$N_{\cPZ}^{\text{misid}}/N_{\cPZ}^{\text{proxy}}$ constitutes the
transfer factor, where
$N_{\cPZ}^{\text{misid}}$ is the number of $\cPZ\to\Pep\Pem$
events in the control sample with an \Pep or \Pem misidentified as a
photon, while $N_{\cPZ}^{\text{proxy}}$ is the number of
$\cPZ\to\Pep\Pem$ events in the control sample with the proxy
condition applied.
The control sample for the \Pe\Pgg\ channel is
taken from the data set collected with the same diphoton trigger as the
signal candidates, while the sample for the \Pgm\Pgg\ channel is from a
data set based on a trigger that requires at least one isolated
electron.
To ensure that the samples dominantly consist of $\cPZ\to\Pep\Pem$
decays, events with one high-purity electron object (``tag'') are
selected from the respective data sets. The
photon candidate and the electron-proxy object are called
``probes''. For each sample of probe candidates, a fit is performed to
the invariant mass
distribution of the tag-probe system to extract the number of
$\cPZ\to\Pep\Pem$ decays.

The ``tag-and-probe'' method described above~\cite{Khachatryan:2010xn}
is executed in bins of three variables: the transverse momentum of the probe
object ($\PT^{\text{probe}}$), the track multiplicity of the primary
vertex ($N_{\text{track}}$), and the number of reconstructed
interaction vertices in the event ($N_{\text{vtx}}$). To account for
the correlations in the distributions of the three variables, the
dependence of the transfer factor on these quantities is modelled by a
three-dimensional parametric function, which is then used to assign an
event-by-event weight to the proxy sample.  The transfer factor is a
decreasing function of $\PT^{\text{probe}}$ and $N_{\text{track}}$,
and an increasing function of $N_{\text{vtx}}$. For a median
value of $N_{\text{vtx}}$, its value varies from 0.04 for events with low
$\PT^{\text{probe}}$ and low $N_{\text{track}}$, to 0.007 for high
$\PT^{\text{probe}}$ and high $N_{\text{track}}$. The relative
uncertainty in the transfer factor is typically of order 10\%, which is
dominated by systematic uncertainties such as those arising from the
tag-and-probe fitting procedure and the parametrization of the transfer
factor. The dependence of the transfer factor on $N_{\text{vtx}}$ is
approximately linear, with a value that changes from about 0.005 to 0.012 for
$N_{\text{vtx}}$ values between 5 (low pileup) and 25 (high pileup).

The estimation of the jet-to-photon misidentification background
follows the same procedure of defining a proxy sample and scaling it
with the transfer factor.
The proxy sample for events with a jet-to-photon
misidentification is constructed by inverting the requirements on the
variable describing the ECAL cluster shape ($\sigma_{\eta \eta}$ in
Ref.~\cite{Khachatryan:2015hwa}) and on one of the isolation variables in
the photon selection.
The transfer factor for the hadronic-proxy sample is determined
through an assessment of
the fraction of events with jet-to-photon misidentification
among the photon candidates. This fraction is denoted as the ``hadron
fraction''.
This measurement is
performed in a low-\MET control sample from a fit to the distribution of
$\sigma_{\eta\eta}$ based on two templates, one representing pure
photons and one modelling the events with jet-to-photon misidentification. The fit is
performed with photon candidates in muon-photon events, where a very
small contamination of misidentified electrons is expected in the photon
sample.  The pure photon template is obtained from $\cPZ\to\Pgm
\Pgm \Pgg$ data by tagging two muons and requiring the three-body
\Pgm\Pgm\Pgg\ invariant mass to be consistent with the \cPZ\ boson
mass. The template that models events with jet-to-photon misidentification is obtained
by inverting the isolation requirement on the signal-photon candidates.
The hadron fraction is measured in \pt\ bins of the photon candidate and
decreases, in general, as a function of \PT. In the \Pe\Pgg\ channel, its
value varies from $0.25 \pm 0.03$ at $\PT=40\GeV$ to $0.08 \pm 0.02$ at
$\PT=120\GeV$. In the \Pgm\Pgg\ channel, the corresponding values are
$0.30 \pm 0.03$ and $0.09 \pm 0.02$. The uncertainties are dominantly
due to possible mismodelling of the fit templates. The small difference
in the hadron fraction of the photon candidates between the \Pe\Pgg\ and
\Pgm\Pgg\ channels originates from small differences in trigger
requirements on the photon object between the diphoton and muon-photon
triggers used to select the \Pe\Pgg\ and \Pgm\Pgg\ data sets.

The \PT\ distribution of the photon objects is multiplied by the hadron
fractions determined as described above. In
the \Pe\Pgg\ channel, the estimated \PT\ distribution of misidentified
electrons is subtracted first. The resulting distribution provides the
estimate of the \PT\ shape for the events with jet-to-photon misidentification. Rather
than forming the ratio of this distribution with the \PT\ distribution of
the proxy sample, both distributions are parameterized individually by
simple analytic functions. The ratio of these two parameterizations
constitutes the transfer factor for the jet-to-photon misidentification.

\subsection{Misidentified-lepton and electroweak backgrounds}

The misidentified-lepton and electroweak (EWK) backgrounds
are evaluated together, as described below.
A misidentified lepton is defined as a reconstructed lepton that does
not arise directly from \PW\ or \cPZ\ boson decays, nor from \Pgt~decays
that originate from a \PW\ or \cPZ\ boson.
Misidentified-lepton events arise primarily from the decay of
heavy-flavour quarks and from hadrons misidentified as leptons, with
other sources such as decays-in-flight constituting a much smaller contribution.
Events where both the lepton and photon are misidentified,
which constitute up to 30\% of the total misidentified-photon
background, are already accounted for by the procedure
described in Section~\ref{subsec:fake_photons}.
The SM electroweak background is dominated by events
with V\Pgg\ (V=\PW, \cPZ) production. In particular, \PW\Pgg\ events have
the same signature as signal events: an energetic
photon, a lepton, and significant \MET. The EWK background
includes rare multiboson events and events with \ttgamma\ production,
which we collectively refer to as the ``rare EWK" background.
Rare EWK events provide only a minor contribution to
the overall background but are relevant in the high-\MET
signal region.

Similar to the determination of the misidentified-photon background,
proxy samples are formed and scaled by transfer factors
to estimate the contribution of misidentified
leptons to the signal region. Each event in the proxy sample contains at
least one candidate photon and at least one misidentified-lepton proxy, but no
candidate lepton.  Proxy objects that model misidentified leptons are
selected by inverting the isolation condition in the lepton
selection. For electrons, the track-cluster matching requirements are
also inverted to further enrich the proxy sample with hadronic
objects. The calculation of the transfer factor used to evaluate the
misidentified-lepton background is described below.

The modelling of the EWK background is based on MC
simulation. Samples of \PW\Pgg, \cPZ\Pgg, \ttgamma, and
\PW\PW\Pgg\ events, listed in the order of decreasing overall
background contributions, are generated with up to two additional partons using the
{\MADGRAPH}\,5 1.3~\cite{Alwall:2014hca} event generator. The
{\PYTHIA}\,6.4~\cite{Sjostrand:2006za} program is used to describe the
parton shower and hadronization.  The \PYTHIA program is further
used to generate samples of \PW\cPZ\Pgg\ events, which produce
an even smaller background contribution
than \PW\PW\Pgg\ events. All samples use the
CTEQ6L1~\cite{Pumplin:2002vw}
parton distribution functions (PDF).
Simulated minimum-bias events are overlaid on the main
hard-scattering events to simulate pileup. The generated particles are
processed through the full CMS detector simulation framework based on
the \GEANTfour~\cite{Agostinelli:2002hh} package, and are subjected to the
same event
selection procedure as the data, including the
trigger requirements. Differences between simulation and data in the
pileup profile, trigger efficiency, and object identification efficiency
are corrected by reweighting the MC events by factors that
lie within a few percent of unity. The \ttgamma, \PW\PW\Pgg, and \PW
\cPZ\Pgg\ samples are normalized to the integrated luminosity of the
data using cross sections calculated with the event generators,
which are valid to leading order (LO) in quantum chromodynamics.  For the
\ttgamma\ sample, a next-to-leading order (NLO)-to-LO scale factor of
2.0~\cite{Melnikov:2011ta} is applied to the cross section to account
for higher-order contributions.

For the V\Pgg\ background, calculated cross sections are used to fix the
ratio between the \PW\Pgg\ and \cPZ\Pgg\ components, but the overall
normalization of the combined sample is derived from data to mitigate
potential uncertainties in the theoretical calculation. This is
accomplished through a two-component template fit describing the
V\Pgg\ and misidentified-lepton backgrounds.
The templates originate from two background samples obtained using the
event selection criteria for the V\Pgg\ MC sample and for the
misidentified-lepton proxy sample.
Distributions of the
variable $\Delta \phi(\ell, \ptvecmiss)$ from the two background samples
in the control region $40<\MET<70\GeV$ are employed as templates. The
lower bound $\MET=40\GeV$ is applied to reduce the contribution of
\cPZ\Pgg\ events. Expected contributions from the misidentified-photon
background and rare EWK backgrounds are subtracted from the data
distribution before the fit. The fit provides scale factors for the template
histograms that are used directly as transfer factors for the
V\Pgg\ and the misidentified-lepton proxy samples. Besides avoiding
a reliance on the value of the
theoretical V\Pgg\ cross section, which is observed to underestimate the
measured production rate of \PW\Pgg\ events~\cite{Aad:2013izg,Chatrchyan:2013fya}, this method has
the benefit that it does not double count the contributions of
background events with both a misidentified photon and lepton. This
class of events is already accounted for in the misidentified-photon
background sample, as mentioned above.

\begin{figure}[bthp]
  \centering
    \includegraphics[width=0.49\textwidth]{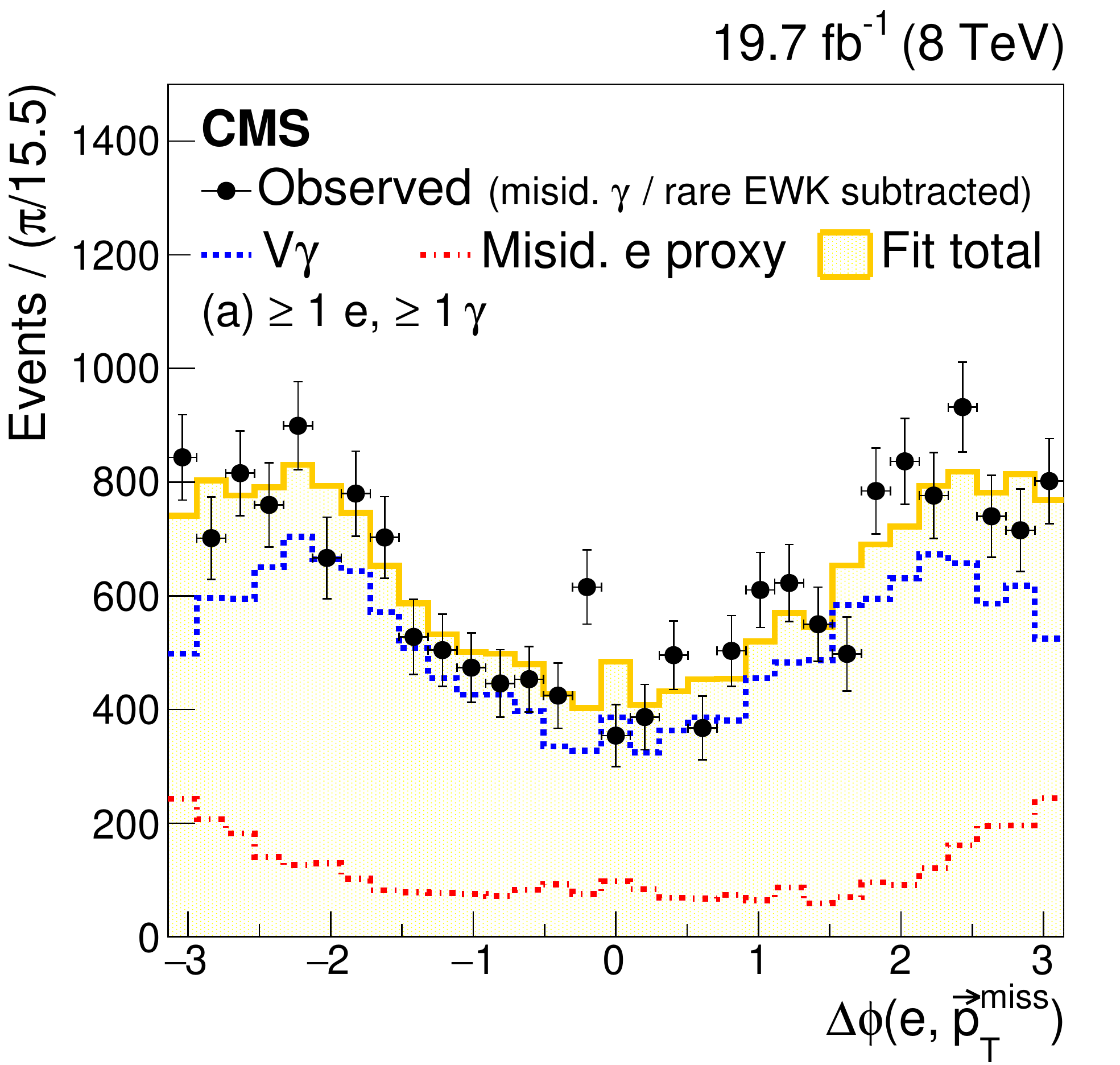}
    \includegraphics[width=0.49\textwidth]{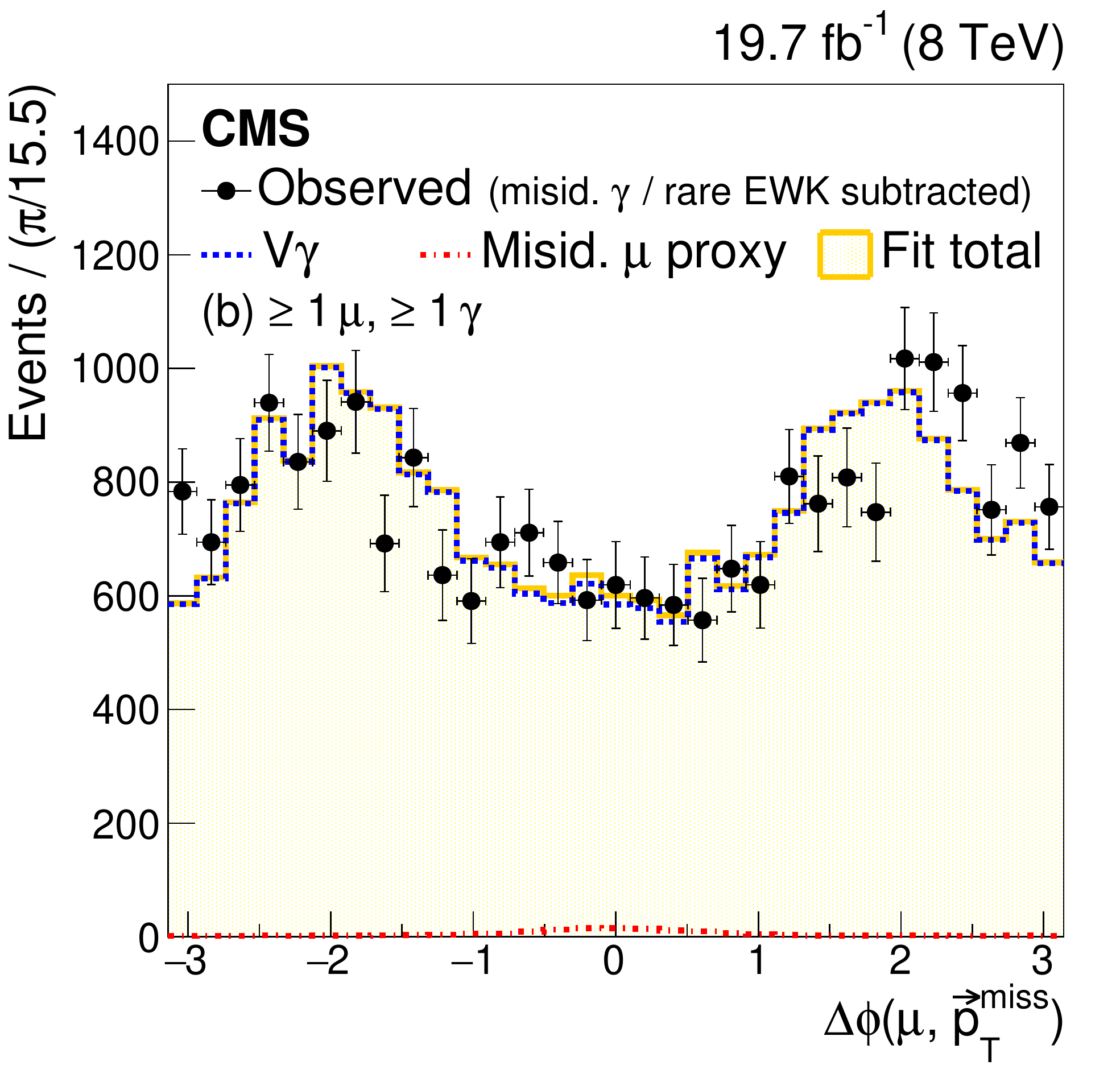}
    \caption{
      Results of the $\Delta\phi(\ell, \ptvecmiss)$ template fit
      for events with $40 < \MET < 70\GeV$, used to determine the
      V\Pgg\ and misidentified-lepton background for the (a) \Pe\Pgg\ and
      (b) \Pgm\Pgg\ channels.
    }
    \label{fig:dPhiFit}
\end{figure}

Figure~\ref{fig:dPhiFit} shows the results of the
template fit, which is performed in the \Pe\Pgg\ and \Pgm\Pgg\ channels
independently.  The resulting scale factors for the V\Pgg\ background in
the \Pe\Pgg\ and \Pgm\Pgg\ channels are $1.59 \pm 0.27$ and $1.47 \pm 0.16$,
respectively, which are similar to each other as expected.
The uncertainties in the scale factors are estimated through toy MC
studies repeating the fit after changing the contributions of the
subtracted misidentified-photon and rare EWK components by their
uncertainties.

\section{Systematic uncertainties}

Table~\ref{tab:systematic_uncertainties} summarizes the sources of
systematic uncertainty for the background predictions and the signal
yields. For each source, the size of the uncertainty is given (in percentage)
relative to the number of events in the corresponding background or
signal sample. For the background, the size of the uncertainty
relative to the total number of background events is also shown.
If the relative uncertainties differ significantly among background
samples because of statistical fluctuations due to the limited number
of events available for the evaluation of the systematic uncertainties, the
range from the minimum to the maximum relative uncertainty is shown.
The dominant experimental uncertainty for the background prediction is
due to the normalization scale
factors applied to the rare EWK and V\Pgg\ samples.  For the rare EWK
sample, a 50\% uncertainty is assigned as a conservative approximation
of the uncertainty in the NLO-to-LO cross section ratio of
\ttgamma\ production, which is the dominant
component in this sample. Also, for the rare EWK sample, we evaluate
the uncertainty due to the luminosity
determination~\cite{CMS:2013gfa}. Normalization uncertainties in the
background estimates of events with misidentified photons or leptons are absorbed in the
uncertainty of the V\Pgg\ scale factor through the uncertainty estimation
in the $\Delta \phi(\ell, \ptvecmiss)$ template fit described
above. Subdominant systematic uncertainties arise from potential mismodelling of
the shapes of the V\Pgg\ and proxy samples for misidentified photons
and leptons.

\begin{table*}[tbp]
  \centering
    \topcaption{
      Summary of systematic uncertainties. The third column gives
      the uncertainty relative to the number of events in the
      corresponding background or signal sample. The fourth column
      shows, for the background terms, the uncertainty relative to the
      total number of background events in the signal region.
    }
    \label{tab:systematic_uncertainties}
    \begin{tabular}{llcc}
      \hline
      \multirow{2}{*}{Name} & \multirow{2}{*}{Sample} & \multicolumn{2}{c}{Relative uncertainty (\%)} \\
      \cline{3-4}
      & & Sample & Total \\
      \hline
        Rare background rate & Rare EWK & 50 & 19 \\
        V\Pgg\ scale factor & V\Pgg & 14 & 6 \\
        Proxy sample shape & Misidentified $\Pgg \& \ell$ & 20--27 & 5 \\
        Trigger and identification efficiency & Rare EWK & 8 & 3 \\
        JES & EWK & 0--6 & 2 \\
        V\Pgg\ shape & V\Pgg & 5 & 2 \\
        Integrated luminosity & Rare EWK & 2.6 & 1 \\
        JER & EWK & 0-2 & 1 \\
        \hline
        JES & Signal & 0-22 & --- \\
        JER & Signal & 0-17 & --- \\
        Trigger and identification efficiency & Signal & 8 & --- \\
        Initial-state radiation & Signal & 0--5 & --- \\
        Integrated luminosity & Signal & 2.6 & --- \\
        \hline
        Renormalization scale and PDF & Signal & 4--41 & --- \\
        \hline
    \end{tabular}

\end{table*}

For simulation-based background estimates,
differences with respect to the data in the jet energy scale (JES)
and resolution (JER) are considered as systematic uncertainties.
The uncertainty associated with the JES is evaluated by varying the
scale by ${\pm}1 \sigma$, where $\sigma$ is the half-width of the
68\% confidence interval around the nominal value, and recalculating the \MET, \MT, and \HT
values in the V\Pgg\ and rare EWK samples, and similarly for the JER term.
The shift in the expected event yield in the signal region
is taken as the estimate of the systematic uncertainty.

Table~\ref{tab:systematic_uncertainties} also lists the systematic uncertainties
considered for the signal MC samples that are used for the interpretation of the result
of this search. The uncertainties due to the JES and JER are evaluated
using the procedure described above. In addition, for the signal samples, uncertainties
in the description of initial-state
radiation as well as the renormalization scale and
PDF~\cite{Kramer:2012bx} are considered.

\section{Results}

\begin{figure*}[btph]
  \centering
  \includegraphics[width=0.4\textwidth]{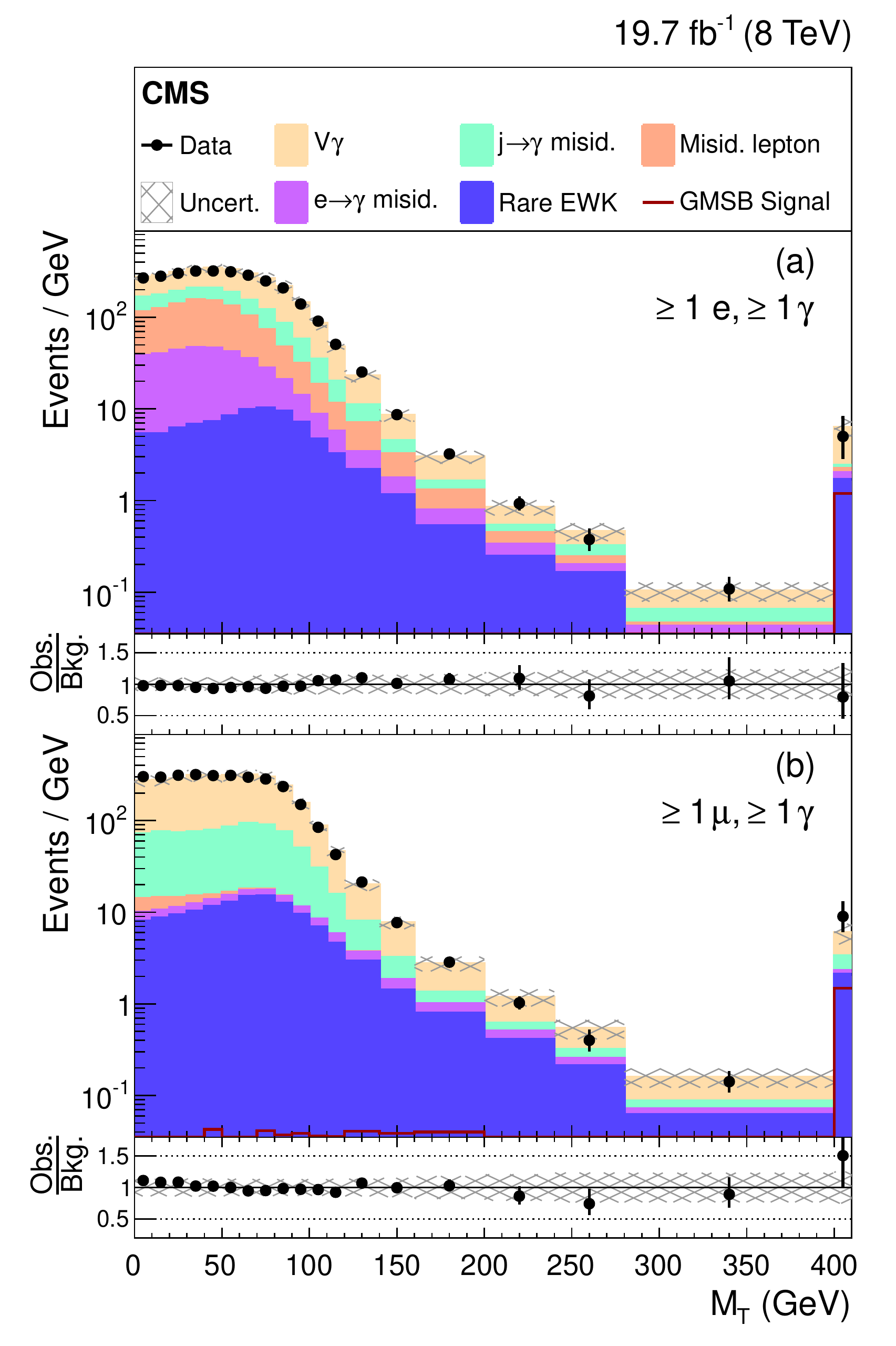}%
  \includegraphics[width=0.4\textwidth]{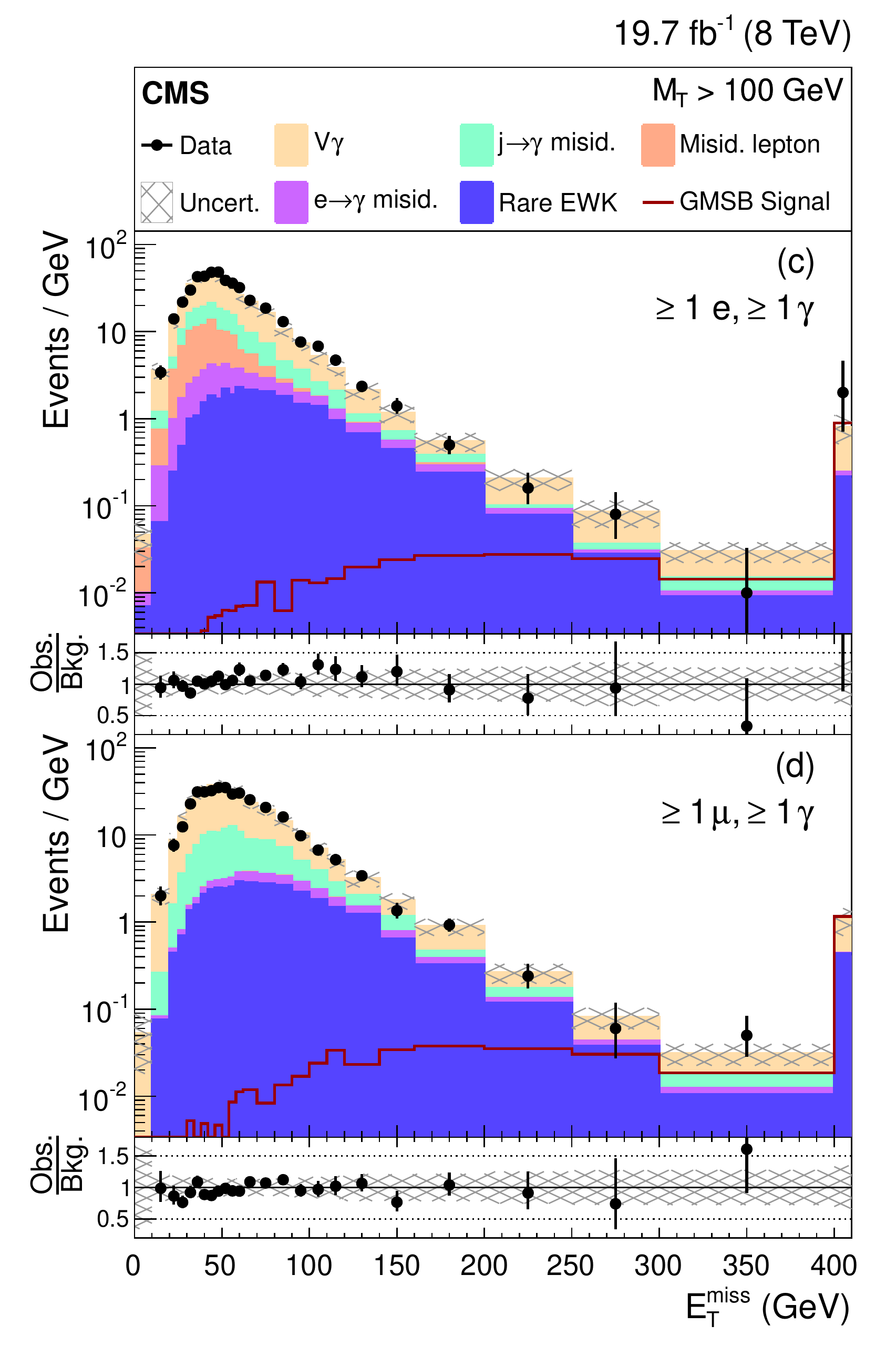}
  \includegraphics[width=0.4\textwidth]{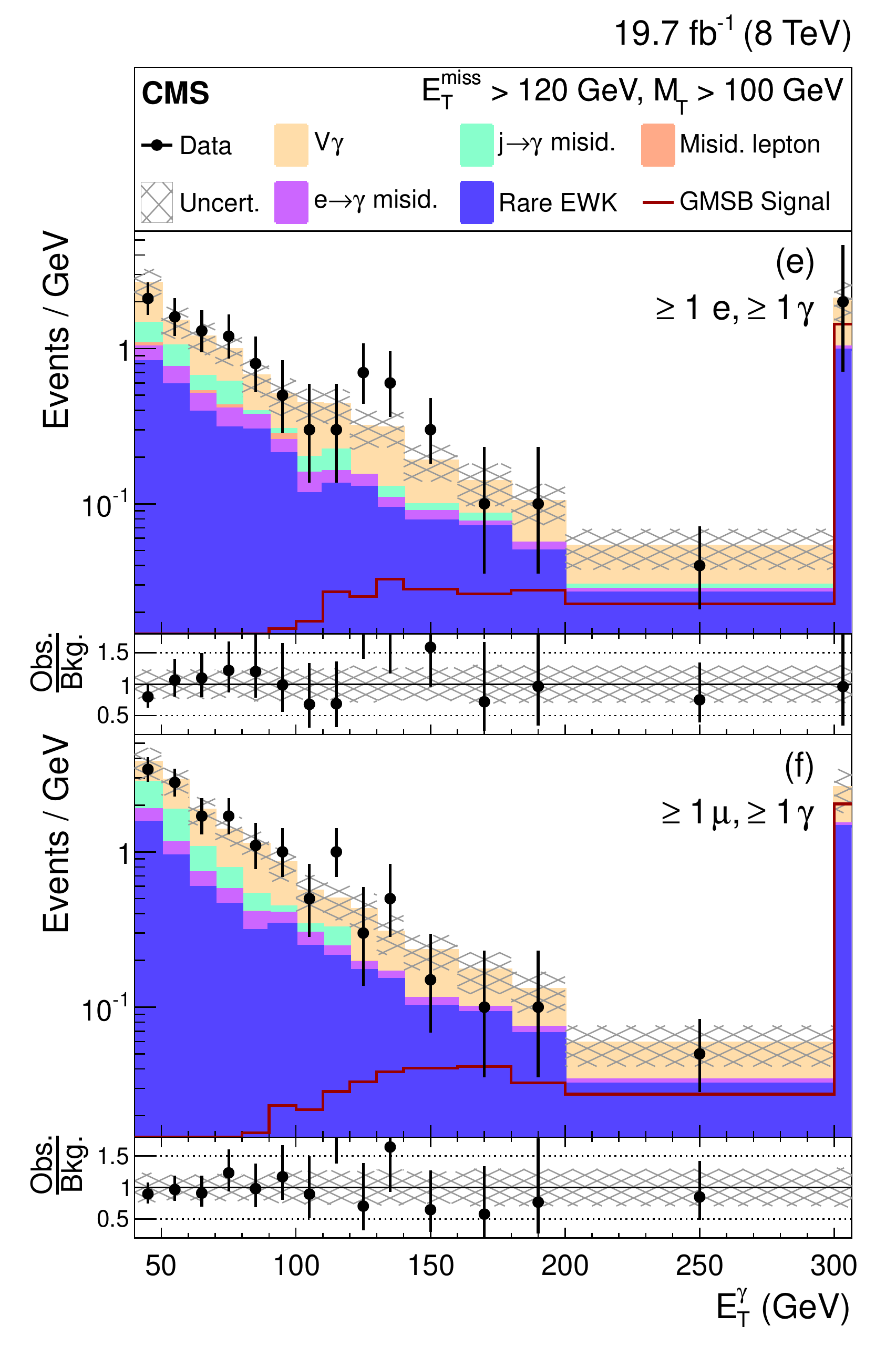}%
  \includegraphics[width=0.4\textwidth]{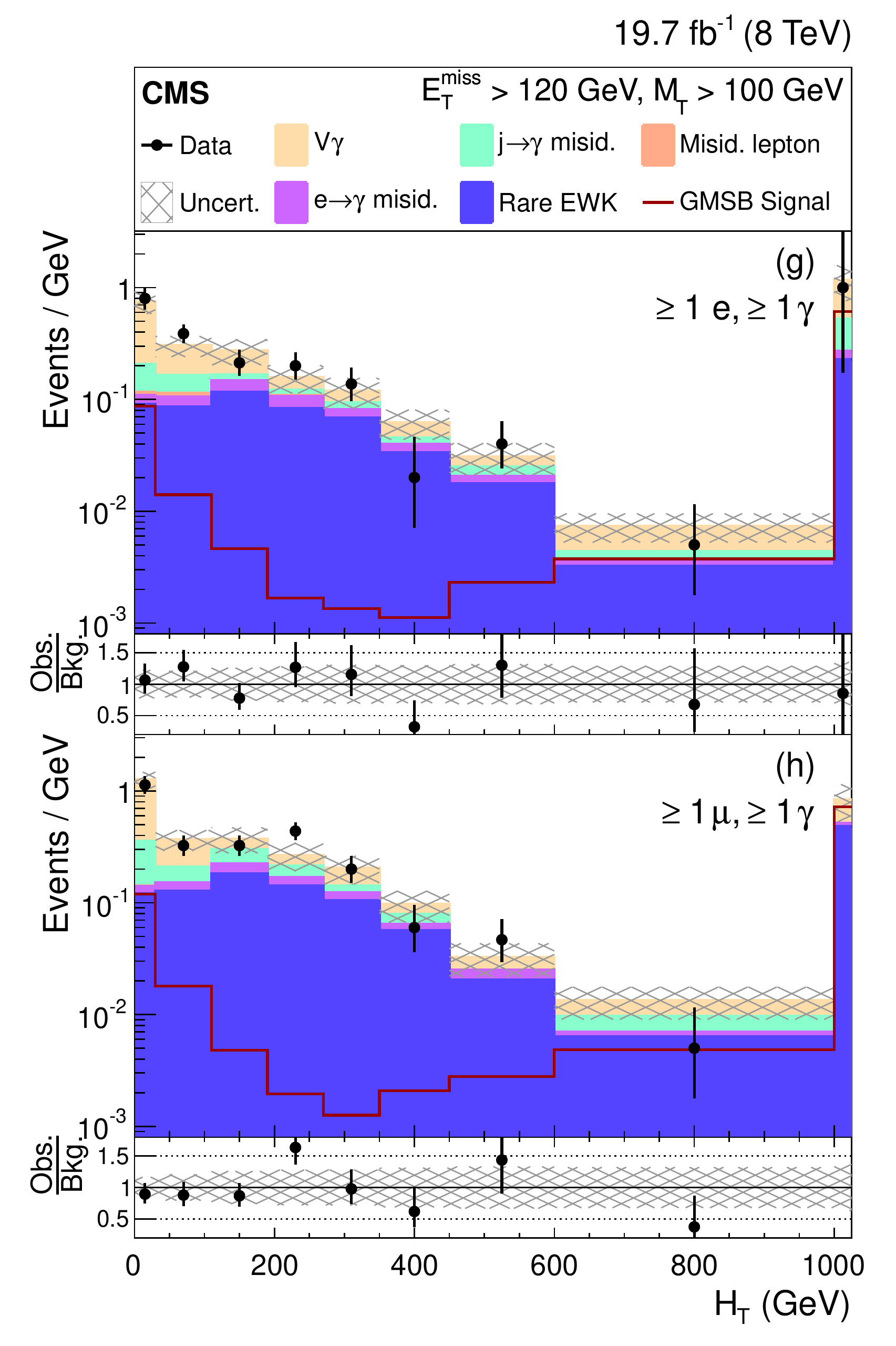}
  \caption{
    Distributions of (a, b) \MT, (c, d) \MET, (e, f) $\ET^{\Pgg}$,
    and (g, h) \HT compared with the stacked
    background expectations for the (a, c, e, g) \Pe\Pgg\ and (b, d,
    f, h) \Pgm\Pgg\ channels.
    See text for details of the event selections applied to these distributions.
    The rightmost bin of each plot shows the overflow, with contents
    that are not normalized by the bin width. Expected signal distributions
    from a GMSB model for a representative
    mass point $(M_{\PSg}, M_{\PSW}) = (915, 405)\GeV$ are overlaid.
    The lower part of each panel shows the ratio of the data to the
    predicted background. The uncertainty bands represent the
    statistical and systematic terms added in quadrature.  }
    \label{fig:stacks}
\end{figure*}

Figure~\ref{fig:stacks} shows the observed distributions of \MT, \MET,
photon \ET ($\ET^{\Pgg}$), and \HT in the \Pe\Pgg\ and \Pgm\Pgg\ channels,
together with the background expectations obtained as described in
Section~\ref{sec:backgrounds}. The
ratio of the observed number of events to the total background
expectation is displayed in the lower part of each panel. The
\MT distribution includes all events that satisfy the event selection
criteria of Table~\ref{tab:cutflow} except for the restrictions on
\MT and \MET. Events in the \MET distribution must additionally
satisfy $\MT > 100\GeV$, and events in the $\ET^{\Pgg}$ and \HT distributions
$\MET > 120\GeV$ and $\MT > 100\GeV$.
The uncertainty bands shown for the background estimates
are the statistical and systematic terms added in quadrature. The
data are seen to agree with the SM prediction within the uncertainties.

\begin{figure*}[bt!h]
  \centering
    \includegraphics[width=0.8\textwidth]{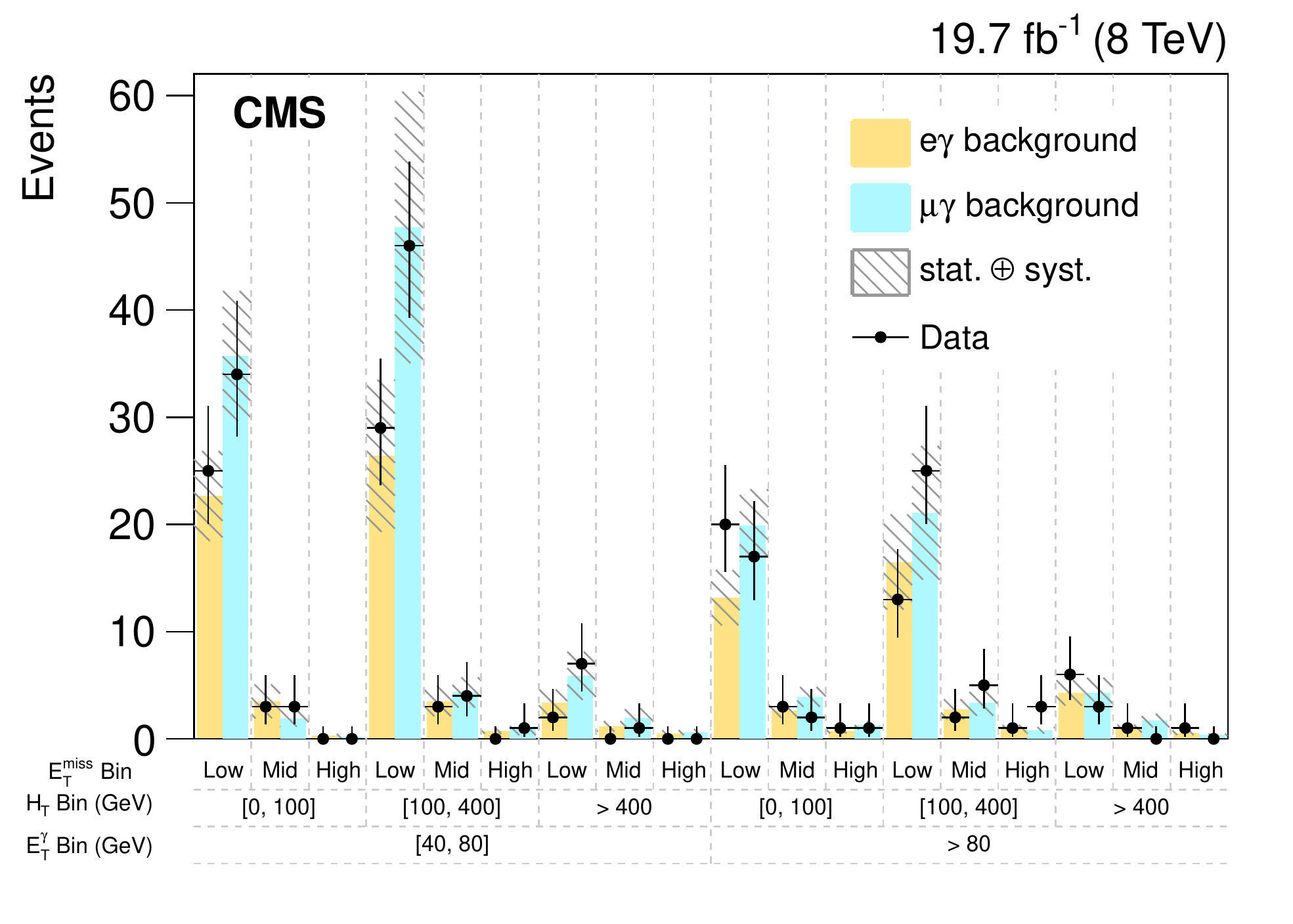}
    \caption{
      Event yields in all signal region bins, compared with the
      combined SM background predictions. The numbers in brackets indicate
      the range of the bins in the signal region variables \MET, \HT,
      and $\ET^{\Pgg}$.  For the \MET bins, Low, Mid, and High
      correspond to the intervals $[120,200]$, $[200,300]$, and
      ${>}300\GeV$, respectively.
    }
    \label{fig:barchart}
\end{figure*}

Figure~\ref{fig:barchart} shows a compilation of event yields compared
to the total background expectations. To enhance the sensitivity to different
SUSY scenarios, the signal region is
further explored in bins of $\ET^{\Pgg}$ ($[40,80]$ and ${>}80\GeV$), \HT
($[0,100]$, $[100,400]$, and ${>}400\GeV$), and \MET (Low, Mid, and High,
corresponding to $[120,200]$, $[200,300]$, and ${>}300\GeV$,
respectively). The data are found to be consistent with the
background prediction in all regions. Thus no significant excess of events beyond the SM expectation is
observed.

\section{Interpretation}
\label{sec:interpretation}

The results of the search are interpreted in terms of cross section
upper limits on a GMSB model and two distinct simplified models. For
each parameter point of the three models, a large number of
hard-scattering simulation events are generated. These events are
processed with a detailed fast simulation of the CMS detector
response~\cite{CMS:2010spa}. A large number of minimum-bias interactions
are superimposed 
on the hard-scattering process in order to reproduce the pileup
profile observed in data. The event selection applied to the simulated signal events is
identical to that applied to data, including the trigger
requirements. The resulting event yields are weighted by correction
factors to account for selection efficiency differences between data and
simulation.

For each mass point of the signal models, a 95\% confidence level (CL)
cross section upper limit is obtained utilizing the ``LHC-style''
$\mathrm{CL}_{\mathrm{s}}$
prescription~\cite{Junk:1999kv,Read:2002hq,CMS-NOTE-2011-005}, which
calculates frequentist $\mathrm{CL}_{\mathrm{s}}$ limits using the
one-sided profiled likelihood as a test statistic. The SM background
prediction, signal expectation, and observed number of events in each
signal-region bin of the \Pe\Pgg\ and \Pgm\Pgg\ channels as shown in
Fig.~\ref{fig:barchart} are combined into one statistical
interpretation, turning the analysis into a multichannel counting
experiment.

\subsection{Interpretation in a GMSB model}

A GMSB model with wino co-NLSPs~\cite{Ruderman:2011vv}, which contains
both electroweak and strong production as the primary SUSY production
process, is examined. All SUSY particles except for the gluino and
winos are considered in the limit of very large mass values such that
they do not participate in the interactions. In this limit, the
lightest chargino and neutralino become purely
wino-like. There is no restriction on the decays, but the gluino always
undergoes at least a three-body decay and the charged wino decays to a
\PW\ boson and the gravitino. The neutral wino decays to a gravitino and
either a photon or a \cPZ\ boson, with branching fractions dictated by
the weak mixing angle and the wino mass.  In the generated scans, the
gluino mass ($M_{\PSg}$) ranges from 715 to 1415\GeV in
50\GeV steps, and the wino mass ($M_{\PSW}$) from 205\GeV to
$[M_{\PSg} - 10\GeV]$, also in 50\GeV steps.

The SUSY particle spectra and branching fractions are determined using
the {\textsc{SuSpect}}\,2.41~\cite{Djouadi:2002ze} and
{\textsc{sdecay}}\,1.3~\cite{Muhlleitner:2003vg} programs.  {\PYTHIA}\,6.4 is
employed for the SUSY particle generation, decays, and the subsequent
parton showers. The cross section for each mass point is determined to
NLO accuracy in quantum chromodynamics using the
{\PROSPINO}\,2.1~\cite{Beenakker:1996ed} program. This cross section
result, along with its uncertainty, is used to derive 95\% CL
exclusion limits on the SUSY particle masses.

\begin{figure}[btp]
  \centering
    \includegraphics[width=\cmsFigWidth]{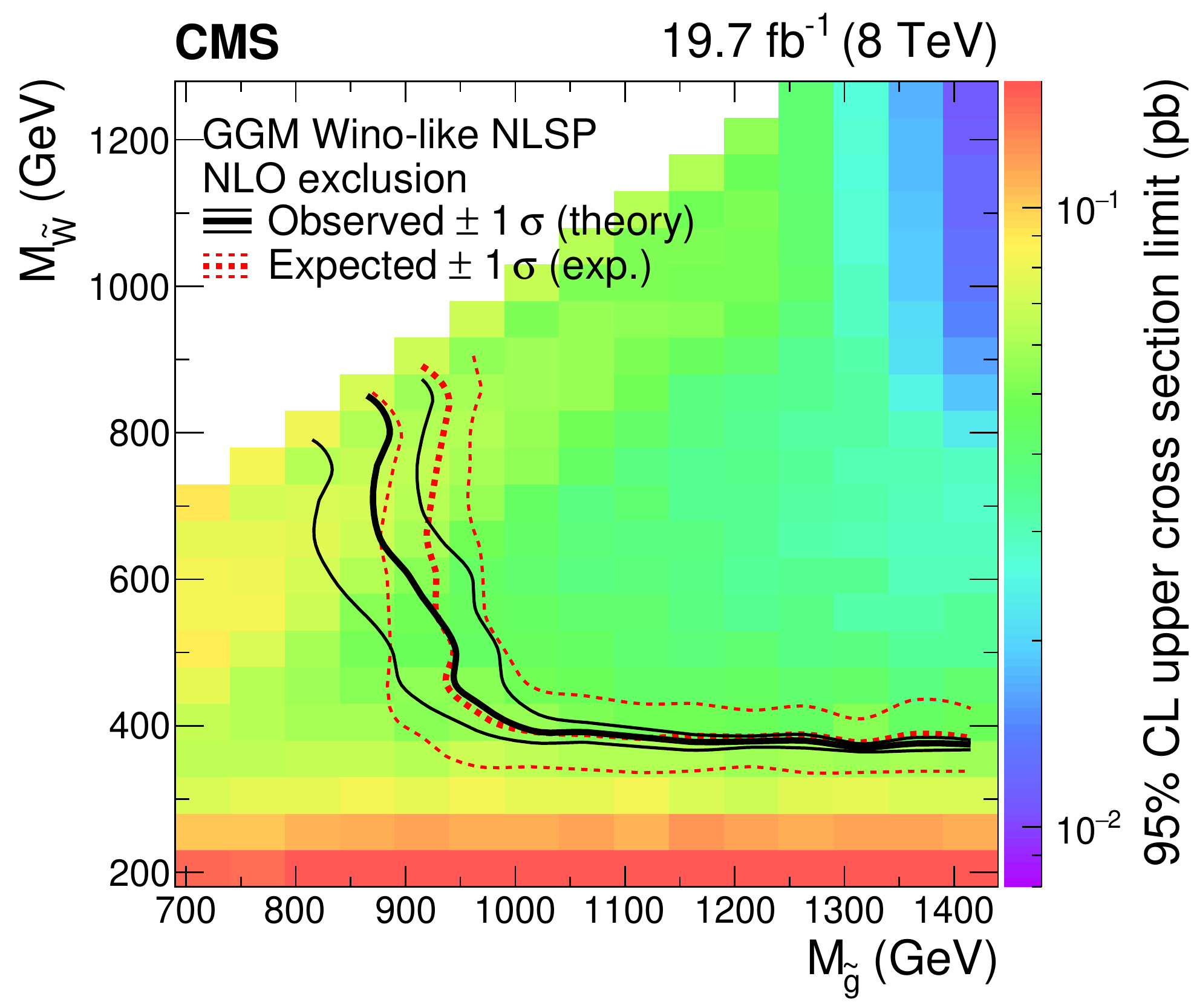}
    \caption{
      Observed and expected 95\% confidence level upper limits on the
      cross section and corresponding exclusion limits for the GMSB model.
    }
    \label{fig:gmsb_limit}
\end{figure}

Figure~\ref{fig:gmsb_limit} shows the observed 95\% CL cross
section upper limits with the exclusion contours overlaid. In the figure, the
dashed curves are the median and $\pm 1$ standard deviation ($\sigma$) 
expected exclusion
contours assuming the nominal cross sections for the signal model.
The solid curves represent the observed exclusion with the signal cross
sections at the nominal and ${\pm}1 \sigma$ values. All mass points on the bottom left side of the contours are
excluded.
Note that the approximate one standard deviation discrepancy between the
expected and observed exclusion contours toward larger values of
$M_{\PSW}$ is due to an upward fluctuation observed in a single bin of
the signal region with $\ET^{\Pgg}>80\GeV$, $\MET>300\GeV$, and an
intermediate bin in \HT from 100 to 400\GeV. 

In this inclusive GMSB model, electroweak production will always take place
when the wino is light, independent of the gluino mass. Thus the
exclusion curve becomes horizontal for
$M_{\PSW}\approx360\GeV$. Note that for this and for all other
instances in this paper where a numerical result is quoted for a mass
limit, the result is based on the theoretical prediction for the cross
section minus its $1 \sigma$ uncertainty. The expected distributions for
signal in Fig.~\ref{fig:stacks} correspond to the GMSB model for the mass
point $(M_{\PSg}, M_{\PSW}) = (915, 405)\GeV$. This mass point has
competing contributions from strong and electroweak SUSY production and
exhibits a non-trivial behavior in \HT as can be seen 
in Fig.~\ref{fig:stacks}(g) and (h).

\subsection{Interpretation in simplified models}

In a simplified model~\cite{Alwall:2008ag, Alves:2011sq,Alves:2011wf}, a
limited set of hypothetical particles and decay chains are introduced to
describe a given topological signature such as the $\ell\Pgg$ final state
studied in this analysis. The production and decay amplitudes of these
particles are parameterized in terms of the particle
masses.

The two simplified models considered are denoted the TChiWg and T5Wg models.
The TChiWg model is initiated by the direct production of hypothetical
particles \PSGcpm\ and \PSGcz, whose decays are restricted to
$\PW^{\pm}\PXXSG$ and $\Pgg\PXXSG$, respectively. The gravitino
\PXXSG\ is nearly massless as in GMSB models. Thus, this model can be
identified with electroweak production in the GMSB wino co-NLSP model,
depicted by the diagram in Fig.~\ref{fig:SMS_cartoon}(a), differing
only in the decay branching fractions. The particles \PSGcpm\ and
\PSGcz\ are therefore identified with gauginos in the remainder of this paper.
A mass
range of $100 \leq M_{\PSGc} \leq 800\GeV$ is considered, where $M_{\PSGc}$ is the
degenerate mass of the gauginos. The generation of
events for the T5Wg model, corresponding to the diagram in
Fig.~\ref{fig:SMS_cartoon}(b), starts with the pair production of
gluinos. The two gluinos undergo three-body
decays $\PSg\to\Pq\Paq\PSGcpm$ and
$\PSg\to\Pq\Paq\PSGcz$, followed by the decays of the \PSGcpm\ and
\PSGcz\ as discussed above. The T5Wg samples are generated in a mass
region $700 \leq M_{\PSg} \leq 1400\GeV$ and $25\GeV \leq M_{\PSGc} \leq
[M_{\PSg} - 25\GeV]$. No other non-SM
particle is involved in either model.

Events for both models are generated with the
{\MADGRAPH}\,5 1.3 program, with up to two final-state partons
in addition to the hard interaction. The events are then
interfaced to {\PYTHIA}\,6.4, which is used to describe the SUSY decay
chains and parton showers. The neutralino-chargino
and the gluino pair production cross sections are calculated to NLO and NLO+NLL
(next-to-leading logarithm) accuracy~\cite{Kramer:2012bx}, respectively,
and used to derive 95\% exclusion limits.

\begin{figure*}[btp]
  \centering
    \includegraphics[width=0.46\textwidth]{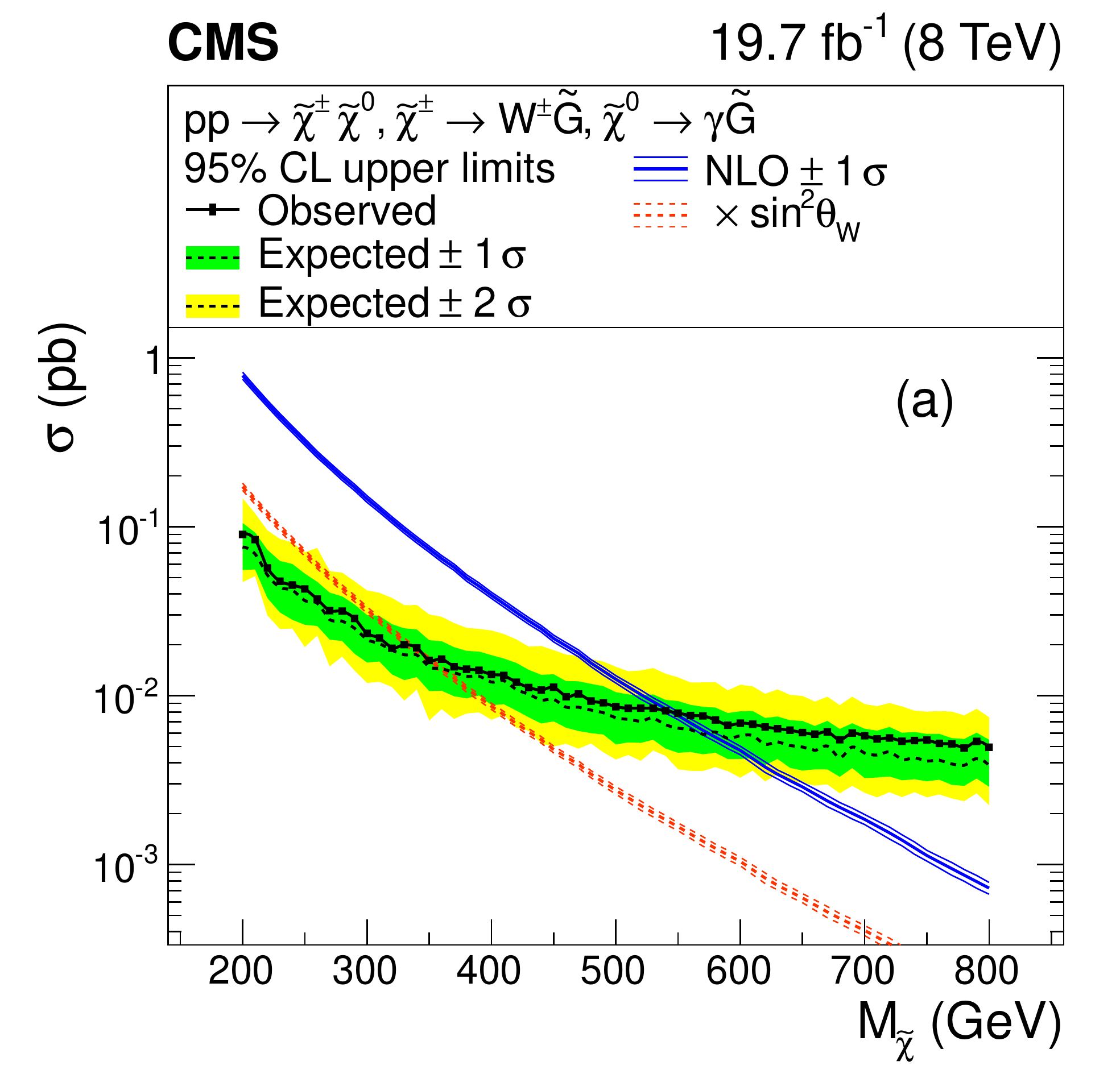}
    \includegraphics[width=0.52\textwidth]{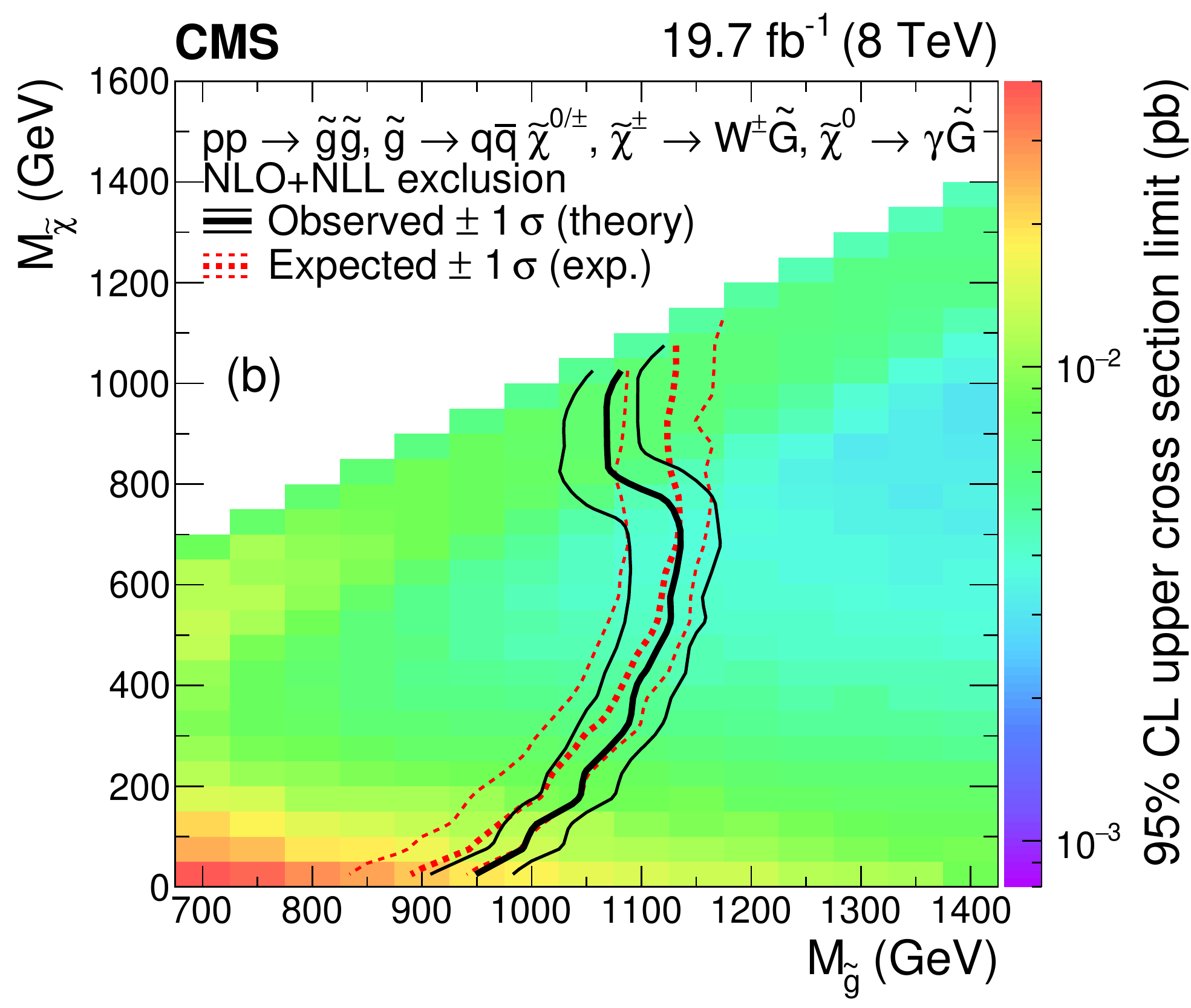}
    \caption{
      Observed and expected 95\% confidence level upper limits on the
      (a) TChiWg and (b) T5Wg simplified models. For the T5Wg model,
      the 95\% confidence level exclusion contour is also shown.
    }
    \label{fig:sms_limit}
\end{figure*}

Figure~\ref{fig:sms_limit}(a) shows the computed 95\%\ CL cross
section upper limit for the TChiWg model as a function of $M_{\PSGc}$,
together with the theoretical cross section. Assuming a 100\% branching
fraction for $\PSGcz \to \Pgg\PXXSG$, the mass range $100 < M_{\PSGc} < 540\GeV$
is excluded, where the lower limit
corresponds to the lowest $M_{\PSGc}$ included in the scan. Assuming a more physically motivated branching fraction
through a rescaling of the theoretical cross section by the weak mixing angle,
the exclusion range is $100 < M_{\PSGc} < 340\GeV$. The latter result is
similar to the limit $M_{\PSW} < 360\GeV$ obtained from the GMSB model with a
wino-like NLSP.

The production cross section of the T5Wg model is determined solely by
$M_{\PSg}$. Nevertheless, the
$M_{\PSg} - M_{\PSGc}$ mass difference affects the \HT and
\MET spectra, resulting in nontrivial exclusion-limit contours in the
$M_{\PSGc}$-$M_{\PSg}$ plane. The 95\% CL cross section upper
 limits and exclusion contours for the T5Wg model are shown in Fig.~\ref{fig:sms_limit}(b).
For $M_{\PSGc} > 200\GeV$, pair production of gluinos is excluded for gluino masses below
1\TeV. For $500 < M_{\PSGc} < 700\GeV$, gluinos below approximately 1.1\TeV are excluded.

\section{Summary}

This paper presents a search for the anomalous production of
events with a photon, an electron or muon, and large missing
transverse momentum produced in proton-proton collisions at
$\sqrt{s}=8\TeV$. The data are examined in bins of the photon
transverse energy, the magnitude of the missing transverse momentum, and \HT,
the scalar sum of jet energies.
The standard model background is evaluated primarily using
control samples in the data, with simulation used to evaluate
backgrounds from electroweak processes.  No excess of events
above the standard model expectation is observed.  The results
of the search are interpreted as 95\% confidence
level upper limits on the production of new-physics events in
the context of a gauge-mediated supersymmetry breaking (GMSB) model.
The GMSB model is excluded for wino masses below 360\GeV.
Results are also interpreted in the context of two simplified
models inspired by GMSB, denoted TChiWg and T5Wg.
The TChiWg model is
excluded for gaugino masses below 540\GeV or,
if the cross sections are scaled by the weak mixing angle,
below 340\GeV.  The T5Wg model with gaugino mass above 200\GeV is excluded for gluino masses
below 1\TeV.

\begin{acknowledgments}
We congratulate our colleagues in the CERN accelerator departments for the excellent performance of the LHC and thank the technical and administrative staffs at CERN and at other CMS institutes for their contributions to the success of the CMS effort. In addition, we gratefully acknowledge the computing centers and personnel of the Worldwide LHC Computing Grid for delivering so effectively the computing infrastructure essential to our analyses. Finally, we acknowledge the enduring support for the construction and operation of the LHC and the CMS detector provided by the following funding agencies: BMWFW and FWF (Austria); FNRS and FWO (Belgium); CNPq, CAPES, FAPERJ, and FAPESP (Brazil); MES (Bulgaria); CERN; CAS, MoST, and NSFC (China); COLCIENCIAS (Colombia); MSES and CSF (Croatia); RPF (Cyprus); MoER, ERC IUT and ERDF (Estonia); Academy of Finland, MEC, and HIP (Finland); CEA and CNRS/IN2P3 (France); BMBF, DFG, and HGF (Germany); GSRT (Greece); OTKA and NIH (Hungary); DAE and DST (India); IPM (Iran); SFI (Ireland); INFN (Italy); MSIP and NRF (Republic of Korea); LAS (Lithuania); MOE and UM (Malaysia); CINVESTAV, CONACYT, SEP, and UASLP-FAI (Mexico); MBIE (New Zealand); PAEC (Pakistan); MSHE and NSC (Poland); FCT (Portugal); JINR (Dubna); MON, RosAtom, RAS and RFBR (Russia); MESTD (Serbia); SEIDI and CPAN (Spain); Swiss Funding Agencies (Switzerland); MST (Taipei); ThEPCenter, IPST, STAR and NSTDA (Thailand); TUBITAK and TAEK (Turkey); NASU and SFFR (Ukraine); STFC (United Kingdom); DOE and NSF (USA).

Individuals have received support from the Marie-Curie program and the European Research Council and EPLANET (European Union); the Leventis Foundation; the A. P. Sloan Foundation; the Alexander von Humboldt Foundation; the Belgian Federal Science Policy Office; the Fonds pour la Formation \`a la Recherche dans l'Industrie et dans l'Agriculture (FRIA-Belgium); the Agentschap voor Innovatie door Wetenschap en Technologie (IWT-Belgium); the Ministry of Education, Youth and Sports (MEYS) of the Czech Republic; the Council of Science and Industrial Research, India; the HOMING PLUS program of the Foundation for Polish Science, cofinanced from European Union, Regional Development Fund; the Compagnia di San Paolo (Torino); the Consorzio per la Fisica (Trieste); MIUR project 20108T4XTM (Italy); the Thalis and Aristeia programs cofinanced by EU-ESF and the Greek NSRF; the National Priorities Research Program by Qatar National Research Fund; the Rachadapisek Sompot Fund for Postdoctoral Fellowship, Chulalongkorn University (Thailand); and the Welch Foundation.
\end{acknowledgments}

\bibliography{auto_generated}

\cleardoublepage \appendix\section{The CMS Collaboration \label{app:collab}}\begin{sloppypar}\hyphenpenalty=5000\widowpenalty=500\clubpenalty=5000\textbf{Yerevan Physics Institute,  Yerevan,  Armenia}\\*[0pt]
V.~Khachatryan, A.M.~Sirunyan, A.~Tumasyan
\vskip\cmsinstskip
\textbf{Institut f\"{u}r Hochenergiephysik der OeAW,  Wien,  Austria}\\*[0pt]
W.~Adam, E.~Asilar, T.~Bergauer, J.~Brandstetter, E.~Brondolin, M.~Dragicevic, J.~Er\"{o}, M.~Flechl, M.~Friedl, R.~Fr\"{u}hwirth\cmsAuthorMark{1}, V.M.~Ghete, C.~Hartl, N.~H\"{o}rmann, J.~Hrubec, M.~Jeitler\cmsAuthorMark{1}, V.~Kn\"{u}nz, A.~K\"{o}nig, M.~Krammer\cmsAuthorMark{1}, I.~Kr\"{a}tschmer, D.~Liko, T.~Matsushita, I.~Mikulec, D.~Rabady\cmsAuthorMark{2}, B.~Rahbaran, H.~Rohringer, J.~Schieck\cmsAuthorMark{1}, R.~Sch\"{o}fbeck, J.~Strauss, W.~Treberer-Treberspurg, W.~Waltenberger, C.-E.~Wulz\cmsAuthorMark{1}
% ...
\vskip\cmsinstskip
% ...
1:~~Also at Vienna University of Technology, Vienna, Austria\\
2:~~Also at CERN, European Organization for Nuclear Research, Geneva, Switzerland\\
% ...

\end{sloppypar}
\end{document}